\documentstyle[12pt]{article}
\newcommand{\be}{\begin{eqnarray}}
\newcommand{\ee}{\end{eqnarray}}
\newcommand{\half}{{\textstyle\frac{1}{2}}}
\newcommand{\Nc}{N_{\rm C}}
\newcommand{\partialslash}{\partial\hspace{-.5em}/\hspace{.15em}}
\newcommand{\ppi}{p_{\rm i}}
\newcommand{\ppf}{p_{\rm f}}
\begin{document}
\rightline{UNITUE-THEP-7/94}
\rightline{hep-ph/9404301}
\rightline{April 1994}
\vspace{1cm}
\begin{center}
\begin{large}
{\bf Off--shell pion electromagnetic form factor from a gauge--invariant
Nambu--Jona-Lasinio model} \\
\end{large}
\vspace{2cm}
{\bf C. Weiss}$^{\rm 1, 2}$ \\
\vspace{0.75cm}
{\em Institut f\"ur Theoretische Physik \\
Universit\"at T\"ubingen \\
Auf der Morgenstelle 14 \\
D--72076 T\"ubingen, Germany}
\end{center}
\vspace{1cm}
\begin{abstract}
\noindent
The off--shell electromagnetic vertex function of pions and kaons is
studied in a bosonized Nam\-bu--Jona-Lasinio model with a gauge--invariant
proper--time cutoff. The slope of the pion form factor with respect to the
pion 4--momentum is equal to the on--shell pion charge radius in the chiral
limit. The off--shell slope of the $K^0$ form factor is zero, that of the
$K^\pm$ about 15\% smaller than that of the pion. We compare with results
of a recent calculation in chiral perturbation theory.
\end{abstract}
\vfill
\rule{5cm}{.15mm}
\\
\noindent
{\footnotesize $^{\rm 1}$ Supported by a research fellowship of the
Deutsche Forschungsgemeinschaft (DFG)} \\
{\footnotesize $^{\rm 2}$ E-mail: weiss@mailserv.zdv.uni-tuebingen.de} \\
\eject
%
%
Electromagnetic interactions of pions play an important role in low--energy
hadronic and nuclear physics. The on--shell electromagnetic form factor of
the charged pion has been measured rather accurately in $\pi e$--scattering
and $e^+ e^-$--annihilation \cite{amendolia_86}. This function has also
been calculated in a variety of dynamical models of the pion, such as the
vector dominance model \cite{sakurai_60}, chiral perturbation theory
\cite{gasser_85}, or models of the Nambu--Jona--Lasinio--type, which
describe the dynamical breaking of chiral symmetry at quark level
\cite{meissner_88}. However, in more complicated processes the
electromagnetically interacting pion is in general not on its mass shell.
For example, pion electroproduction involves the half--off--shell pion
vertex function \cite{naus_89}, while in nuclear electromagnetic processes
with pion exchange currents generally both legs of the pion vertex function
are off--shell \cite{riska_87}. The analysis of such processes requires
knowledge of the electromagnetic vertex function for arbitrary pion
4--momenta.
\par
Recently, Rudy {\em et al.}\ have calculated the off--shell electromagnetic
form factor of the pion in chiral perturbation theory \cite{rudy_94}. For
off--shell pion momenta the vertex function receives a contribution from a
tree--level term of order $O(p^4 )$, which vanishes on--shell by virtue of
the equation of motion of the pion field. This term brings in an unknown
parameter, unrelated to the on--shell pion charge radius, $L_9$, which
leaves the off--shell behavior of the vertex function undetermined at this
level. In contrast, a picture in which the pion is considered as a bound
state in a chirally invariant quark model assigns a definite off--shell
behavior to the pion vertex function.
\par
In this letter, we study the off--shell pseudoscalar meson electromagnetic
vertex function in a bosonized Nambu--Jona--Lasinio (NJL) model
\cite{volkov_83,ebert_86}, defined with a gauge invariant proper--time
cutoff \cite{schwinger_51}. This finite, chirally and gauge--invariant
model action provides an ideal framework for the study of pion
electromagnetic properties. We consider the full momentum dependence of the
meson propagator and vertex function and do not restrict ourselves to a
gradient expansion of the quark loop. Our method is an extension of the
approach recently employed to study diquark electromagnetic form factors
inside baryons \cite{weiss_93}. A general expression for the vertex
function is obtained, which is then analyzed by expanding around the
on--shell point and studying the slopes of the form factors as functions of
the meson mass. The general structure of our results is in agreement with
the one found in chiral perturbation theory \cite{rudy_94}. Moreover, we
obtain a specific prediction for the off--shell behavior of the pion form
factor. In particular, in the chiral limit, the off--shell slope of the
form factor is found to be equal to the on--shell pion charge radius.
Finally, we discuss the role of $SU(3)$--symmetry breaking in the slopes of
the $K^\pm$-- and $K^0$--form factors.
\par
It is well--known in field theory that the off--shell behavior of
correlation functions depends on the choice of interpolating field. Only
matrix elements between asymptotic states are protected by the equivalence
theorem \cite{borchers_60}. Thus, individual results for off--shell
correlation functions should be regarded as building blocks in describing
more complicated processes like {\em e.g.}\ pion Compton scattering or
electroproduction. Nevertheless, it is instructive to compare the
predictions for the off--shell behavior of the pion vertex function in
different but related approaches. Furthermore, considering the difficulty
of a unified field--theoretic description of nuclear processes like
electroproduction, a realistic approach is likely to be patchwork of
different phenomenological descriptions.
\par
We start from the action of the bosonized NJL model in the form
\cite{ebert_86,reinhardt_90}
\be
{\cal S} &=& - i\, {\rm Tr}_\Lambda \log G^{-1}
\, - \, \frac{\Nc}{4g} \int\! d^4 x\, \phi_\alpha \phi^\alpha .
\label{effective_action}
\ee
Here, $g$ is the NJL coupling constant,
$G^{-1} = i \partialslash - m_0 - \phi_\alpha \Lambda^\alpha$ the inverse
quark propagator, with $m_0 = {\rm diag}(m_u , m_d , m_s )$ the current
quark mass matrix, and $\phi_\alpha$ the meson field. The vertices,
$\Lambda^{\alpha}$, are matrices in color, flavor and Dirac spinor space,
$\Lambda^{\alpha} \, = \, 1_{\rm C} (\half\lambda^{a})_{\rm F} O^{\sf a},
\; \alpha = (a, {\sf a})$, where $O^{\sf a} = \{ 1,\, i\gamma_5 ,\,
i\gamma^\mu /\sqrt{2} , \, i\gamma^\mu\gamma_5 / \sqrt{2} \}$,
corresponding to scalar, pseudoscalar, vector and axial vector mesons. The
vacuum value of the scalar meson field defines the constituent quark mass,
$M = {\rm diag}(M_u , M_d , M_s )$ \cite{ebert_86}. To
eq.(\ref{effective_action}) we couple an electromagnetic field by way of
minimal substitution,
$i\partialslash \rightarrow i\partialslash - e Q {\cal A}_\mu \gamma^\mu$,
where $Q = {\rm diag}(\frac{2}{3}, -\frac{1}{3}, -\frac{1}{3})$
is the quark charge matrix, and expand the effective action in the
fluctuating pseudoscalar meson and the electromagnetic field,
\be
{\cal S} &=& {\cal S}_0
\;\, +\;\, \half\int\!\frac{d^4 p}{(2\pi)^4}\;
\phi_{\alpha}(-p)\, ({\cal D}^{-1})^{\alpha\beta}(p)\, \phi_{\beta}(p)
\label{expansion} \\
&+& \int\!\frac{d^4 \ppf}{(2\pi)^4}\,\int\!\frac{d^4 \ppi}{(2\pi)^4}
\;\phi_{\alpha}(-\ppf )\, \phi_\beta (\ppi )\, {\cal A}_\mu (\ppf - \ppi )\,
{\cal F}^{\alpha\beta\mu} (\ppi , \ppf ) \;\, + \;\, \ldots .
\nonumber
\ee
Up to field renormalization, ${\cal D}$ is the meson propagator and
${\cal F}$ the (irreducible) electromagnetic vertex function. For
pseudoscalar mesons ($O^{\sf a} = i\gamma_5$) they are of the form
\be
({\cal D}^{-1})^{\alpha\beta}(p) &=& \Nc \sum_{ij}
\left(\frac{\lambda^a}{2}\right)_{ij} \left(\frac{\lambda^b}{2}\right)_{ji}
\left( -\frac{1}{2g} + I_{ij} (p^2 ) \right) ,
\label{propagator} \\
{\cal F}^{\alpha\beta\mu}(\ppi , \ppf ) &=&
e\Nc \sum_{ij} \left(\frac{\lambda^a}{2}\right)_{ij}
\left(\frac{\lambda^b}{2}\right)_{ji} \left(
F_{ij}(\ppi^2 , \ppf^2 , q^2) (\ppi^\mu + \ppf^\mu )
+ G_{ij}(\ppi^2 , \ppf^2 , q^2) q^\mu \right) , \;\; {}
\label{vertex_function}
\ee
where $q = \ppf - \ppi$. In the absence of flavor mixing, the invariant
functions $I_{ij}, F_{ij}, G_{ij}$ directly correspond to a physical meson
channel if the meson is labeled by its quark flavor content, {\em i.e.},
$F_{ud}$ is the $\pi^+$--vertex function, $F_{us}$ the $K^+$--vertex
function, {\em etc.}.
\par
We define the quark loop in eq.(\ref{effective_action}) using a
gauge--invariant proper--time regularization,
\be
{\rm Re}\,
{\rm Tr}_\Lambda \log G_E^{-1} &=&
\half\int_{\Lambda^{-2}}^\infty\frac{ds}{s}
\,\, {\rm Tr}\,\exp (-s\, G_E^{-1\,\dagger} G_E^{-1}) .
\label{proper_time}
\ee
Here, $G_E^{-1}$ is the quark propagator after continuation to euclidean
space \cite{ebert_86,schwinger_51}. The functions $I_{ij}, F_{ij}, G_{ij}$
are obtained by expanding the proper--time integral,
eq.(\ref{proper_time}), in the fluctuating meson and electromagnetic field
\cite{schwinger_51,weiss_93}. The resulting expressions are then continued
back to Minkowski space. In this way one finds for the meson propagator
\be
I_{ij} (p^2 ) &=& (p^2 - (M_i - M_j)^2 ) A_{ij}(p^2 )
+ M_i^2 B_i + M_j^2 B_j ,
\label{polarization} \\
A_{ij} (p^2 ) &=& \frac{1}{16\pi^2}\, \int_0^1 d\alpha\, \Gamma\left( 0,
\frac{\alpha M_i^2 + (1 - \alpha ) M_j^2 -
\alpha (1 - \alpha ) p^2}{\Lambda^2} \right) ,
\label{A_integral} \\
B_i &=& \frac{1}{16\pi^2} \Gamma\left( -1, \frac{M_i^2}{\Lambda^2} \right) .
\hspace{2em} \nonumber
\ee
For the vertex function, eq.(\ref{vertex_function}), we obtain the general
result
\be
F_{ij}(\ppi^2, \ppf^2, q^2 ) &=& (Q_i - Q_j ) \half \left[
A_{ij}(\ppi^2 ) + A_{ji}(\ppf^2 ) + p_+^2 C^+_{1, ij} + q^2 C^+_{2, ij}
\right]
\label{F_general} \\
&+& (Q_i + Q_j ) \half \left[ p_+^2 C^-_{1, ij} + q^2 C^-_{2, ij} \right] ,
\nonumber \\
G_{ij}(\ppi^2, \ppf^2, q^2 ) &=& (Q_i - Q_j ) \half \left[
- A_{ij}(\ppi^2 ) + A_{ji}(\ppf^2 )
+ p_+^2 C^+_{3, ij} + p_-^2 C^+_{2, ij} \right]
\label{G_general} \\
&+& (Q_i + Q_j ) \half \left[ p_+^2 C^-_{3, ij} + p_-^2 C^-_{2, ij} \right] ,
\nonumber \\
p_+^2 &=& \ppi^2 + \ppf^2 - q^2 - 2(M_i - M_j)^2, \hspace{1.5em}
p_-^2 \; = \; \ppi^2 - \ppf^2 ,
\nonumber \\
C_{k, ij}^\pm &\equiv&
C_{k, ij}^\pm (\ppi^2, \ppf^2, q^2 )
\; =\; \half \left( C_{k, ij} (\ppi^2 , \ppf^2 , q^2 )
\pm C_{k, ji} (\ppi^2 , \ppf^2 , q^2 ) \right) ,
\label{C_integral} \\
C_{k, ij} (\ppi^2 , \ppf^2 , q^2 ) &=&
\frac{1}{16\pi^2}\, \int_0^1 d\alpha\, \int_0^{1 - \alpha}
d\beta\, X_k \, \frac{\exp (-Y^2/\Lambda^2 )}{Y^2} ,
\nonumber \\[.5ex]
Y^2  &=& \alpha M_j^2 + (1 - \alpha) M_i^2 - \alpha (1 - \alpha - \beta)
\ppi^2 - \alpha\beta \ppf^2 - \beta (1 - \alpha - \beta ) q^2 ,
\nonumber \\
X_1 &=& \alpha , \hspace{1.5em} X_2 \; = 1 , \hspace{1.5em}
X_3 \; = 1 - \alpha - 2\beta . \nonumber
\ee
In arriving at eqs.(\ref{F_general}, \ref{G_general}) we have made use of
the symmetry properties
$C_{k, ij} (\ppi^2 , \ppf^2 , q^2 ) = \varepsilon_k
C_{k, ij} (\ppf^2 , \ppi^2 , q^2 ), \; \varepsilon_k = \{ +, +, - \}$.
As a consequence,
$F_{ij} (\ppi^2 , \ppf^2 , q^2 ) = F_{ij} (\ppf^2 , \ppi^2 , q^2 )$ and
$G_{ij} (\ppi^2 , \ppf^2 , q^2 ) = - G_{ij} (\ppf^2 , \ppi^2 , q^2 )$, as
required by time reversal invariance. Note that, for a charged meson
$(\pi^\pm , K^\pm )$, $Q_i - Q_j = \pm 1$, while for a neutral meson
$(\pi^0, K^0, \bar{K}^0 )$, $Q_i - Q_j = 0$, but $Q_i + Q_j \neq 0$. In the
limit of flavor symmetry, $M_i = M_j$, the part of the vertex function
proportional to $Q_i + Q_j$ vanishes identically. We consider here the
isospin limit, $M_u = M_d \neq M_s$. Thus, the $\pi^0$ vertex function is
identically zero, but $K^0 , \bar{K}^0$ have vertex functions proportional
to $\pm (Q_d + Q_s )$.
\par
We note that, as a result of the gauge--invariant definition of the
effective action, eq.(\ref{proper_time}), the meson vertex function and
propagator satisfy the Ward identity
\be
q_\mu \left( F_{ij}(\ppi^2 , \ppf^2 , q^2 ) (\ppi^\mu + \ppf^\mu )
+ G_{ij}(\ppi^2 , \ppf^2 , q^2 ) q^\mu \right)
&=& (Q_i - Q_j ) \left( I_{ij}(\ppf^2 ) - I_{ij}(\ppi^2 ) \right) .
\label{ward_identity}
\ee
\par
On the mass shell, $\ppi^2 = \ppf^2 = m^2$, one has
$C_{3, ij} (m^2 , m^2 , q^2 ) \equiv 0$, and the longitudinal part of the
vertex function vanishes, $G_{ij} (m^2 , m^2 , q^2 ) \equiv 0$. The mass of
the meson bound states of flavor $i\bar\jmath$ is determined as the zero of
the inverse propagator, $I_{ij}(m^2) - \frac{1}{2}g^{-1} = 0$. We
normalize the meson field to unit residue of the propagator at $p^2 = m^2$.
For a given flavor channel this means dividing
$F_{ij}(\ppi^2 , \ppf^2 , q^2)$ and $G_{ij}(\ppi^2 , \ppf^2 , q^2)$ of
eqs.(\ref{propagator}, \ref{vertex_function}) by
\be
Z_{ij}(m^2 ) &=& \frac{\partial}{\partial p^2}\,
I_{ij}(p^2 ) |_{p^2 = m^2} .
\ee
The field renormalization factor, $Z_{ij} (m^2 )$, depends on the meson
mass but is otherwise independent of the field momentum. We discuss in the
following the off--shell behavior of the vertex function subject to this
definition of the meson field. This redefinition corresponds to a
(momentum--independent) finite multiplicative renormalization of the meson
source in the original generating functional of the NJL model\footnote{For
the bosonization of the NJL model including meson source terms, see
\cite{kaschluhn_92}.}. The normalized on--shell meson charge form factor
is then given by $F_{ij}(m^2, m^2, q^2 )/Z_{ij} (m^2 )$.
\par
In order to exhibit the off--shell behavior of the vertex function, and to
make contact with the results of chiral perturbation theory, we expand
eqs.(\ref{F_general}, \ref{G_general}) simultaneously in the momentum
transfer, $q^2$, and the off--shellness, $\ppi^2 - m^2$ and $\ppf^2 - m^2$.
In practice, this is an excellent approximation, as the full expressions
eqs.(\ref{F_general}, \ref{G_general}) are almost linear in these variables
for momenta up to $\sim 0.25 {\rm GeV}^2$. The expansion is of the form
\be
Z_{ij}^{-1} (m^2 ) F_{ij} (\ppi^2 , \ppf^2 , q^2 ) &=& (Q_i - Q_j )
\left[ 1 + r^+_{ij} (m^2 ) q^2 + s^+_{ij} (m^2 )
(\ppi^2 + \ppf^2 - 2 m^2 ) + \ldots\; \right]
\nonumber
\\
&+& (Q_i + Q_j ) \left[ r^-_{ij} (m^2 ) q^2 + s^-_{ij} (m^2 )
(\ppi^2 + \ppf^2 - 2 m^2 ) + \ldots\; \right] ,
\label{F_expansion}
\\
Z_{ij}^{-1} (m^2 ) G_{ij} (\ppi^2 , \ppf^2 , q^2 ) &=& (Q_i - Q_j )
\left[ t^+_{ij} (m^2 ) (\ppi^2 - \ppf^2 ) + \ldots\; \right]
\nonumber
\\
&+& (Q_i + Q_j ) \left[ t^-_{ij} (m^2 ) (\ppi^2 - \ppf^2 ) + \ldots \;
\right] .
\label{G_expansion}
\ee
The slopes, $r^\pm_{ij} (m^2 ), s^\pm_{ij} (m^2 )$ and $t^\pm_{ij} (m^2 )$,
which are functions of the meson mass, are easily obtained as derivatives
of eqs.(\ref{F_general}, \ref{G_general}). Here, $r^\pm_{ij} (m^2 )$ are
related to the (on-shell) meson r.m.s.\ charge radius, while
$s_{ij}^\pm (m^2 )$ are the off--shell slopes of the form factor. From the
Ward identity, eq.(\ref{ward_identity}), it follows that
$t^\pm_{ij} (m^2 ) \equiv r^\pm_{ij} (m^2 )$. Furthermore, using the
symmetry properties of the Feynman parameter integrals,
eq.(\ref{C_integral}), one finds $s^-_{ij} (m^2 ) \equiv 0$ for any flavors
$i, j$ and arbitrary $m^2$.
\par
It is worthwhile to investigate the slopes of the meson form factor in
dependence of the meson mass, {\em i.e.}, on the symmetry--breaking
current quark masses. For the charged pion in the isospin limit,
$M_u = M_d$, the slopes of the meson form factor are
$r_{\pi^+} (m_\pi^2 ) = r^+_{ud} (m_\pi^2 )$ and $s_{\pi^+} (m_\pi^2 ) =
s^+_{ud} (m_\pi^2 )$. In particular, in the chiral limit, $m_\pi^2 = 0$,
the Feynman parameter integrals determining $r_{\pi^+} (0)$ and
$s_{\pi^+}(0)$ can be performed trivially, and one obtains
\be
r_{\pi^+} (0) &=& s_{\pi^+} (0) \; = \; \frac{Z^{-1}_{ud}(0)}{96\pi^2
M_u^2} \exp (-M_u^2 / \Lambda^2 ) \; = \;
\frac{\Nc}{24\pi^2 f_\pi^2} \exp (-M_u^2 / \Lambda^2 ) .
\label{r_0}
\ee
Thus, in the chiral limit the off--shell slope of the pion form factor is
equal to the $q^2$--slope, {\em i.e.}, the charge radius. (In the last
equation we have made use of the fact that, in the chiral limit,
$Z_{ud} (0) = A_{ud} (0) = \frac{1}{4}\Nc^{-1} M_u^{-2} f_\pi^2$.)
\par
In chiral perturbation theory, the off--shell slope of the form factor,
$s_{\pi^+} (m_\pi^2 )$, introduces a new phenomenological parameter
unrelated to the meson charge radius at tree level, $L_9$ \cite{rudy_94}.
In the notation of ref.\ \cite{rudy_94},
$s_{\pi^+} (0) = 16\beta_1 f_\pi^{-2}$, where $\beta_1$ is the parameter
multiplying an $O(p^4 )$--tree level term, which vanishes if the equation
of motion for the pion field holds. From eq.(\ref{r_0}) we find
$\beta_1 = (\Nc /384\pi^2 ) \exp (-M_u^2 / \Lambda^2 )$. In the limit
$\Lambda \rightarrow \infty$, keeping $f_\pi$ fixed, this agrees with the
value quoted by Rudy {\em et al.} \cite{rudy_94}, which they infer by
rewriting the lagrangian obtained from a gradient expansion of the quark
determinant \cite{ebert_86} in a form which exhibits the off--shell term
determining $s_{\pi^+} (0)$. In our treatment this coefficient is obtained
directly, with no need to ``undo'' the equation of motion for the pion
field. Moreover, the direct relation between the off--shell slope and the
charge radius in the chiral limit, eq.(\ref{r_0}), is lost in the other
approach. Crudely speaking, eq.(\ref{r_0}) means that a massless pion and
a photon with $q^2 = 0$ ``see'' the quark loop in the same way, which seems
intuitively plausible.
\par
The coefficients $r_{\pi^+} (m_\pi^2 )$ and $s_{\pi^+} (m_\pi^2 )$ are
shown in fig.1 as functions of the pion mass, which is generated by chiral
symmetry breaking of the form $m_u = m_d \neq 0$. The common parameters are
$M_u = M_d = m_\rho / \sqrt{6} = 315\,{\rm MeV}$, as suggested by the KSFR
relation \cite{ebert_86}, and $\Lambda = 660\,{\rm MeV}$ determined from
fitting $f_\pi = 93\,{\rm MeV}$ at the physical pion mass. (Here, when
changing $m_u = m_d$, we keep $M_{u, d}$ and $\Lambda$ fixed and adjust the
NJL coupling, $g$, according to the gap equation.) With these parameters
the charge radius in the chiral limit is
$\langle r^2 \rangle_{\pi^+} = (0.52\, {\rm fm})^2$. If
$\pi$--$A_1$--mixing is taken into account in fixing the cutoff,
significantly larger values for $\Lambda$ are obtained \cite{ebert_86},
leading to larger values of the pion charge radius; in the limit
$\Lambda\rightarrow\infty$ one finds
$\langle r^2 \rangle_{\pi^+} = (0.58\, {\rm fm})^2$, {\em cf.}\
\cite{tarrach_79}. Fig.1 shows that $r_{\pi^+} (m_\pi^2 )$ and
$s_{\pi^+} (m_\pi^2)$ vary only little from the chiral limit up to the
physical pion mass, so that relation eq.(\ref{r_0}) is well satisfied for
physical pions.
\par
Eq.(\ref{expansion}) takes into account the direct coupling of the photon
to the pion through the quark loop. This is not necessarily in
contradiction to the vector dominance picture, as the mass of the
$\bar qq$--intermediate state is approximately equal to $m_\rho$.
Moreover, the quark core radius, eq.(\ref{r_0}), is reasonably close to the
experimental value, $\langle r^2 \rangle_{\pi^+} = (0.66\, {\rm fm})^2$.
We therefore have reason to believe that the quark loop also accounts for
most of the off--shell behavior of the form factor\footnote{The
contributions of the quark core and of composite pion loops to the pion
charge radius have recently been estimated in the framework of an effective
quark theory \cite{bender_93}.}. Our intention here is to compare with
chiral perturbation theory, which does not include resonance contributions.
The coupling of the photon through the vector mode of
eq.(\ref{effective_action}) should, however, be included when studying the
pion form factor in the timelike region.
\par
For completeness we note that the off--shell behavior of the pion
propagator is governed by the same parameter as that of the vertex
function, {\em i.e.}, one has
$I_{ij} (p^2 ) = Z_{ij} (m^2 ) (p^2 - m^2)
[ 1 \, + \, s^+_{ij}(m^2) (p^2 - m^2 ) \, + \, \ldots ]$ near $p^2 = m^2$.
This result is required in constructing the reducible 3--point Green's
function from the vertex function.
\par
In the neutral kaon vertex function only the part proportional to
$(Q_d + Q_s)$ contributes. Since $s^-_{ds} (m_K^2 ) \equiv 0$, the $K^0$
form factor does not show any off--shell dependence to order
$(\ppi^2 - \ppf^2 - m_K^2 )$, in agreement with the chiral perturbation
theory calculation \cite{rudy_94}. However, this result is completely
independent of the quark masses and the kaon mass and thus not obviously
related to chiral symmetry, as suggested in \cite{rudy_94}.
\par
In evaluating the charged kaon form factors and the $q^2$--slope of the
neutral kaon form factor, one has to take into account effects of
$SU(3)$--symmetry breaking. We consider symmetry breaking of the form
$M_s \neq M_u = M_d$, corresponding to $m_s \neq m_u = m_d = 0$. In doing
so we have to keep in mind that $m_K^2$ depends on $m_s$. Fig.1 shows the
$K^+$ charge radius and off--shell slope,
$r_{K^+}(m_K^2 ) = (Q_u - Q_s) r^+_{us} (m_K^2 ) + (Q_u + Q_s) r^-_{us}
(m_K^2 )$, and
$s_{K^+}(m_K^2 ) = (Q_u - Q_s) s^+_{us} (m_K^2 ) + (Q_u + Q_s) s^-_{us}
(m_K^2 )$, as well as the $K^0$ charge radius,
$r_{K^0}(m_K^2 ) = (Q_d + Q_s) r^-_{ds} (m_K^2 )$, as a function of
$m_K^2$. (Here, $M_{u, d}$ and $\Lambda$ are as above, $g$ is kept fixed,
while $M_s$ and $m_K$ vary in dependence on $m_s$.)  At the physical kaon
mass, the $K^+$ charge radius is roughly 10\%, the off--shell slope 15\%
smaller than the value for the pion, while the $K^0$ charge radius is
around 20\% of the pion charge radius. These ratios are rather insensitive
to the cutoff and the constituent quark mass. One may also evaluate
$r_{K^+}(m_K^2 ), s_{K^+}(m_K^2)$ and $r_{K^0}(m_K^2 )$ by expanding to
leading order in $M_s - M_u$. The slopes change of order $M_s - M_u \sim
m_K^2$ when going away from the chiral limit. However, the first--order
approximation to $SU(3)$--breaking does not provide a reliable quantitative
estimate of the kaon radii, since the non--linear terms are large in this
model, {\em cf.}\ fig.1. We note that the present approach takes into
account $SU(3)$--symmetry breaking effects to all orders. This is different
from chiral perturbation theory, where symmetry breaking is governed by
terms of order $\log m_0$ originating from pion and kaon loops. The
relations $r_{K^+} = r_{\pi^+} + r_{K^0}$ of \cite{gasser_85} and
$s_{K^+} = s_{\pi^+}$ of \cite{rudy_94} for physical pion and kaon masses
only hold up to ``ordinary'' chiral symmetry breaking effects of order
$m_s$, {\em cf.}\ the comments in \cite{gasser_85}. As fig.1 shows, they
are satisfied at the 10\% --level in the NJL model.
\par
In summary, we have discussed the off--shell electromagnetic vertex
function for pions and kaons in a gauge--invariant version of the bosonized
NJL model. The off--shell slope of the form factor coincides with the
charge radius in the chiral limit. The pion vertex function thus exhibits a
high degree of symmetry; it is essentially characterized by one parameter
--- the pion charge radius. Modifications for finite pion masses are small.
We note that the smooth nature of the chiral limit in this approach is
essential in obtaining that result. It would be interesting to see if
relation eq.(\ref{r_0}) is realized also in other dynamical models of the
pion, {\em i.e.}, to what extent it depends on the particular choice of
interpolating pion field. The general pattern of the off--shell behavior of
the form factors agrees with the one obtained in chiral perturbation theory
at the level to be expected.
\par
The effective meson action derived from the NJL model,
eq.(\ref{effective_action}), has been generalized to include baryon fields
as composites of diquark and quark fields \cite{reinhardt_90,kaschluhn_92}.
The resulting meson--baryon theory allows in principle to calculate also
the pion--nucleon part of the pion electroproduction amplitude, based on
the same quark dynamics. In this framework the off--shell effects in the
pion form factor, the pion--nucleon vertex and the pion propagator could be
incorporated consistently, {\em i.e.}, with the same interpolating pion
field.  Such an approach, which requires at least approximate knowledge of
the baryon diquark--quark wave function, remains an interesting
possibility.
\newpage
\newpage
\noindent {\Large\bf Figure caption}
\\[.4cm]
Fig.1: The $q^2$--slope and the off--shell slope of the electromagnetic
form factor of pions and kaons, as a function of the meson mass. All
quantities are expressed in units of the slope of the pion form factor in
the chiral limit, $r_{\pi^+} (0)$, {\em cf.}\ eq.(\ref{r_0}). Common
parameters are $M_u = M_d = 315\,{\rm MeV}$ and $\Lambda = 660\,{\rm MeV}$.
{\em Thin solid and dashed line}: $r_{\pi^+} (m_\pi^2 )/r_{\pi^+} (0)$ and
$s_{\pi^+} (m_\pi^2 )/r_{\pi^+} (0)$, with $m_\pi$ generated by
isospin--symmetric chiral symmetry breaking of the form
$m_u = m_d \neq m_s = 0$. {\em Fat solid and dashed line}:
$r_{K^+} (m_K^2 )/r_{\pi^+} (0)$ and $s_{K^+} (m_K^2 )/r_{\pi^+} (0)$;
{\em fat dot--dashed line}: $r_{K^0} (m_K^2 )/r_{\pi^+} (0) + 1$ (note the
shift by $+1$; $r_{K^0} (0) = 0$).  Here, $m_K$ and $M_s$ are driven by
$SU(3)$--symmetry breaking of the form $m_s \neq m_u = m_d = 0$. Note that
the pion and kaon masses in this plot are determined by different types of
explicit chiral symmetry breaking. The physical pion and kaon masses are
indicated by arrows.
\end{document}